\documentclass[aps,prl,twocolumn,showpacs,superscriptaddress,footinbib]{revtex4-1} 
\usepackage{amssymb}
\usepackage{amsmath}
\usepackage[colorlinks=true,linkcolor=blue,citecolor=blue, urlcolor=blue]{hyperref}   
\usepackage{tikz}
\usepackage{graphicx}
\usepackage[utf8]{inputenc}
\definecolor{uibred}{RGB}{167, 38, 47}

\def\Fig#1{Fig.~\ref{#1}}

\def\p{\mathbf{p}}

\begin{document}

\title{Prescaling and Far-from-Equilibrium Hydrodynamics in the Quark-Gluon Plasma} 

\author{Aleksas Mazeliauskas}
\email[]{a.mazeliauskas@thphys.uni-heidelberg.de}
\affiliation{Institut f\"{u}r Theoretische Physik, Universit\"{a}t Heidelberg, Philosophenweg 16, D-69120 Heidelberg, Germany}

\author{J\"{u}rgen Berges}
\email[]{berges@thphys.uni-heidelberg.de}
\affiliation{Institut f\"{u}r Theoretische Physik, Universit\"{a}t Heidelberg, Philosophenweg 16, D-69120 Heidelberg, Germany}

\begin{abstract}
Prescaling is a far-from-equilibrium phenomenon which describes the rapid establishment of a universal scaling form of distributions 
much before the universal values of their scaling exponents are realized. We consider the example of the spatio-temporal 
evolution of the quark-gluon plasma explored in heavy-ion collisions at sufficiently high energies. Solving QCD kinetic theory with elastic 
and inelastic processes, we demonstrate that the gluon and quark distributions very quickly adapt a self-similar scaling form,
which is independent of initial condition details and system parameters. The dynamics in the prescaling regime is then fully encoded in a
few time-dependent scaling exponents, whose slow evolution gives rise to far-from-equilibrium hydrodynamic behavior. 
\end{abstract}

\pacs{11.10.Wx, 
     }
   
\maketitle

{\it Introduction.} 
Universal scaling phenomena play an important role in our understanding of the thermalization process in quantum many-body systems. Topical applications range from heavy-ion collisions~\cite{Baier:2000sb,Berges:2013eia,Kurkela:2014tea} to quenches in ultracold quantum gases~\cite{Orioli:2015dxa, Mikheev:2018adp,
Prufer:2018hto,Erne:2018gmz}. Starting far-from-equilibrium, these systems exhibit transiently a nonthermal fixed point regime where the time evolution of characteristic quantities becomes self-similar. Consequently, details about initial conditions and underlying system parameters become irrelevant in this regime, and the nonequilibrium dynamics is encoded in universal scaling exponents and functions~\cite{Micha:2002ey,Berges:2013eia,Schlichting:2012es,Berges:2013fga,Orioli:2015dxa,York:2014wja,Berges:2015ixa}.  

In heavy-ion collisions at sufficiently high energies, where the gauge coupling is small due to asymptotic freedom~\cite{Gross:1973id, Politzer:1973fx}, the time evolution of gluons (g) and quarks (q) is described by distribution functions $f_{g,q}(p_\perp, p_z,\tau)$. Since the 
system is longitudinally expanding, the distributions depend on transverse ($p_\perp$) and longitudinal momenta ($p_z$), and on proper time ($\tau$)~\cite{Bjorken:1982qr,Baym:1984np}. In the scaling regime the gluon distribution obeys 
\begin{equation}
f_g(p_\perp, p_z,\tau) \stackrel{\mathrm{scaling}}{=}  \tau^{\alpha} f_S\big(  \tau^{\beta}p_\perp, \tau^{\gamma}p_z\big), 
\label{eq:scaling}
\end{equation}
with dimensionless $\tau \rightarrow \tau/\tau_{\text{ref}}$ and $p_{\perp,z} \to p_{\perp,z}/Q_s$ in terms of some (arbitrary) time $\tau_\text{ref}$ and characteristic momentum scale $Q_s$. The exponents $\alpha$, $\beta$ and $\gamma$ are universal and the nonthermal fixed-point distribution $f_S$ is universal up to normalizations~\cite{Berges:2013eia}, which has been established numerically using classical-statistical lattice simulations~\cite{Berges:2013fga}. The exponents are expected to be $\alpha_\text{BMSS}=-2/3$, $\beta_\text{BMSS}=0$ and $\gamma_\text{BMSS}=1/3$ according to the first stage of the `bottom up' thermalization scenario~\cite{Baier:2000sb} based on number conserving and small-angle scatterings, or $\alpha_\text{BD}=-3/4$, $\beta_\text{BD}=0$ and $\gamma_\text{BD}=1/4$ in a variant of `bottom-up' including the effects of plasma instabilities~\cite{Bodeker:2005nv}. 

In this Letter, we compute the evolution of the quark-gluon plasma approaching the nonthermal fixed-point using leading-order QCD kinetic theory~\cite{Arnold:2002zm}. Since this state-of-the-art description involves elastic and inelastic processes, conservation of particle number is not built in, and no small-angle approximation is assumed~\cite{Kurkela:2018oqw, Kurkela:2018xxd}. 

We establish that the far-from-equilibrium dynamics according to leading-order QCD kinetic theory exhibits self-similar scaling. Comparing to the dynamics with elastic scatterings only, the softer momentum regions are efficiently populated by collinear radiation processes, which is seen to improve the universal scaling behavior of the distributions. 

Most remarkably, we find that much before the scaling (\ref{eq:scaling}) with universal exponents is established, the evolution is already governed by 
the fixed-point distribution $f_S$ as
\begin{equation}
f_g(p_\perp, p_z,\tau) \stackrel{\mathrm{prescaling}}{=}  \tau^{\alpha(\tau)} f_S\big( \tau^{ \beta(\tau)}p_\perp, \tau^{\gamma(\tau)}p_z\big), 
\label{eq:prescaling}
\end{equation}
with non-universal {\it time-dependent} exponents $\alpha(\tau)$, $\beta(\tau)$ and $\gamma(\tau)$. This represents a dramatic reduction in complexity already at this early stage: 
The entire evolution in this prescaling regime is encoded in the time dependence of a few slowly evolving exponents, and we point out the relation to 
hydrodynamic behavior far from equilibrium.

The phenomenon of prescaling describes the rapid establishment of universal nonequilibrium results for certain quantities ($f_S$), though others still deviate from their universal values ($\alpha, \beta, \gamma$). This has to be distinguished from standard corrections due to finite size/time scaling behavior, from which asymptotic universal values are inferred without taking the infinite volume/time limit. Coined by Wetterich based on the notion of partial fixed points~\cite{Wetterich:1981ir,Aarts:2000wi}, prescaling has recently been explored in the context of scaling violations in the short-distance behavior of correlation functions for Bose gases~\cite{Schmied:2018upn}. 

{\it QCD kinetic theory.} We employ the leading order QCD kinetic theory of Ref.~\cite{Arnold:2002zm} to evolve the 
gluon and quark distributions by
\begin{align}
\!\partial_\tau f_{g,q}(\p,\tau)\! -\! \frac{p_z}{\tau}\partial_{p_z} f_{g,q}(\p,\tau)= 
-\mathcal{C}_{g,q}^{2\leftrightarrow2}[f]\!-\mathcal{C}_{g,q}^{1\leftrightarrow2}[f],\!\label{bolz}
\end{align}
using the numerical setup developed in
 Refs.~\cite{Kurkela:2018oqw, Kurkela:2018xxd}.  
Here $\mathcal{C}_{g,q}^{2\leftrightarrow2}[f]$ represents the collision integrals for leading-order elastic scatterings in the coupling $\alpha_s \equiv g^2/(4\pi)$. This involves scatterings $gg\leftrightarrow gg$, $qq\leftrightarrow qq$, $gq\leftrightarrow gq$ as well as particle conversion $gg\leftrightarrow q\bar q$ processes. To this order we also include number changing processes $\mathcal{C}_{g,q}^{1\leftrightarrow2}[f]$ of medium induced collinear gluon radiation $g\leftrightarrow gg$, $q\leftrightarrow qg$ and quark pair production $g\leftrightarrow q\bar q$.
The  effective $1\leftrightarrow 2$ splitting rate is calculated by the resummation of multiple interactions with the medium and includes the Landau-Pomeranchuk-Migdal suppression of collinear radiation~\cite{Landau:1953gr,Landau:1953um,Migdal:1955nv,Migdal:1956tc}.
The soft momentum exchange is regulated by isotropic screening~\cite{Kurkela:2018oqw, Kurkela:2018xxd}.

\begin{figure}[t!]
	\centering
	\includegraphics[width=0.9\columnwidth]{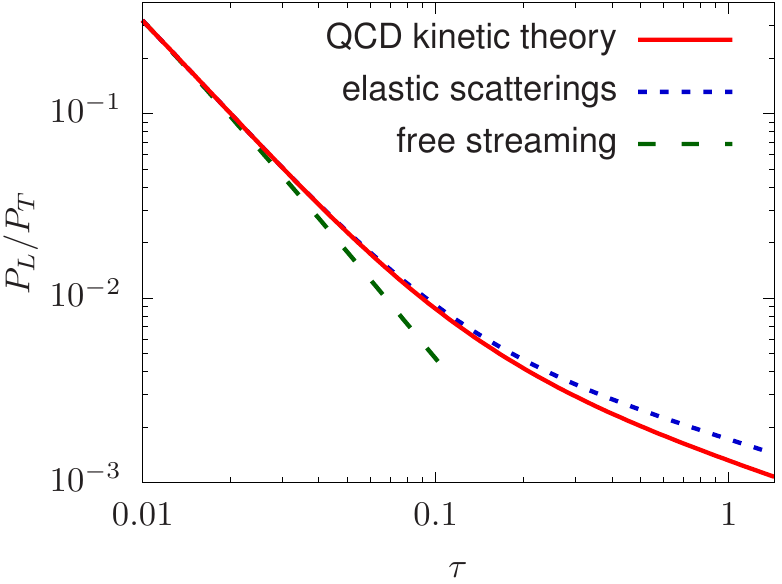}
	\caption{
Pressure anisotropy $P_L/P_T$ for a longitudinally expanding QCD plasma with over-occupied gluon initial state.}
	\label{fig:PLPT}
\end{figure}

{\it From free-streaming to universal scaling.}
We consider the initial distributions $f^0_{g,q}(\p) =A_{g,q}\, \exp\{-(p_\perp^2+\xi^2 p_z^2)/Q_s^2\}$, where $A_q=0.5$ for quarks and  $A_g=\sigma_0/g^2$ for gluons.
For energetic collisions the bosonic gluons are expected to be highly occupied $f(Q_s)\sim 1/g^2$ with a characteristic momentum scale $Q_s$, while fermion occupancies are bounded by Fermi-Dirac statistics~\cite{Lappi:2006fp, Gelis:2010nm}. Here $\sigma_0$ is taken to be $\sigma_0=0.1,0.6$ and $g=10^{-3}$ in view of the range of validity of kinetic theory and in order to make clean comparisons with previous lattice simulations~\footnote{We checked that  prescaling is observable for a wide range of coupling $g=10^{-3}{-}10^{-1}$. For values $g\gtrsim 10^{-2}$ the scaling exponents do not reach the asymptotics of the nonthermal attractor and rather proceed to the subsequent stage of ``bottom-up", but prescaling is still visible.}. The initial anisotropy is controlled by $\xi$, and we employ $\xi = 2$.
Starting at $\tau_0Q_s=70$ and choosing $\tau_\text{ref}Q_s=7000$, we solve the coupled set of kinetic equations (\ref{bolz}) for the gluon and quark distributions numerically~\cite{Kurkela:2018oqw, Kurkela:2018xxd}.

\begin{figure}[t!]
	\centering
   \includegraphics[width=0.88\columnwidth]{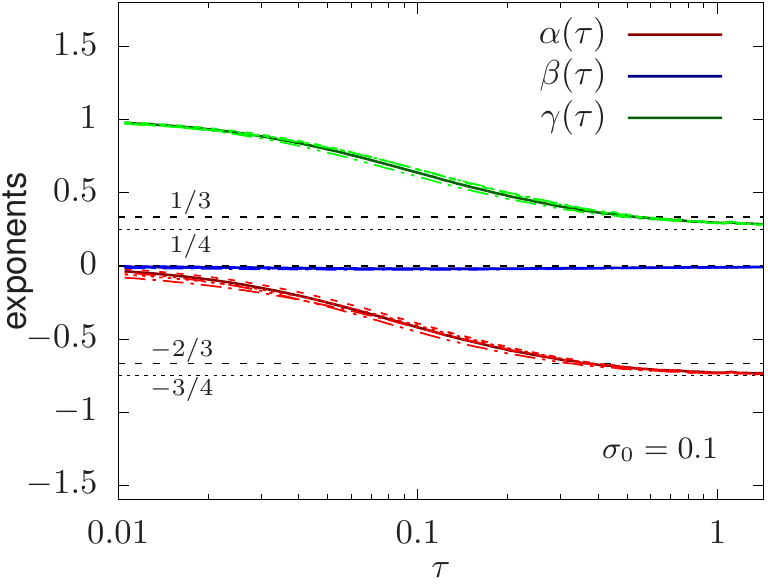}
	\caption{Time-dependent scaling exponents from multiple sets of integral moments for gluon density parameter $\sigma_0=0.1$.
}
	\label{fig:timeexp}\label{fig:plotexponentsqcddivv3}
\end{figure}

\begin{figure}[b!]
	\centering
   \includegraphics[width=0.88\columnwidth]{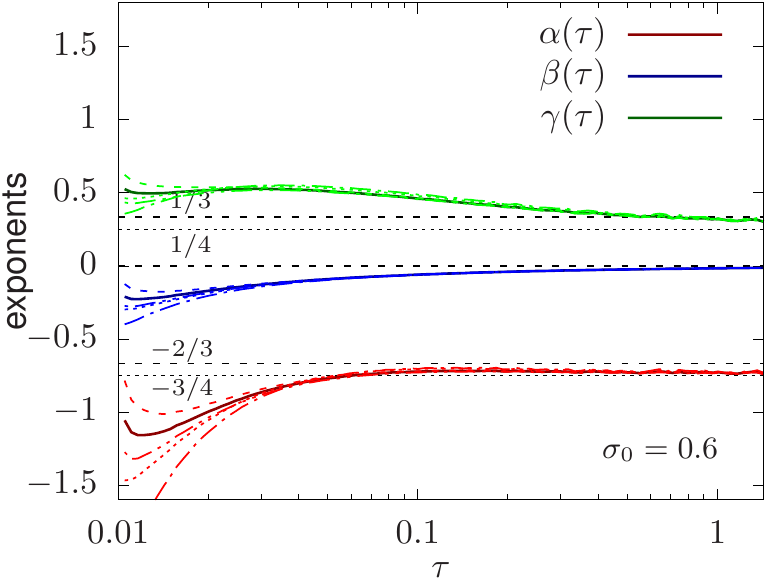}
	\caption{The same as Fig.~\ref{fig:plotexponentsqcddivv3} but for $\sigma_0=0.6$.}
	\label{fig:timeexpb}\label{fig:plotexponentsqcddivv3b}
\end{figure}

In Fig.~\ref{fig:PLPT}, we show the evolution of the pressure anisotropy $P_L/P_T$ (solid curve) as a function of dimensionless time $\tau\rightarrow \tau/\tau_\text{ref}$ for $\sigma_0 = 0.1$. The longitudinal and transverse pressures are defined from the energy-momentum tensor 
\begin{eqnarray}
T^{\mu \nu}(\tau) &=& \int\frac{d^3\p}{(2\pi)^3}\frac{p^\mu p^\nu}{p^0} \big(\nu_g f_g(\p,\tau) +  2N_f \nu_q f_q(\p,\tau)\big) \nonumber\\
&=& \mathrm{diag}\left(e,P_T,P_T,P_L\right),
\label{eq:enmomtensor}
\end{eqnarray}
where $\nu_g = 2(N_c^2-1) = 16$ and $\nu_q = 2 N_c  = 6$ for $N_c = 3$ colors and $N_f = 3$ quark flavors. One observes how the pressure anisotropy starts to deviate from collisionless expansion (dashed line) relevant at earliest times, and bends over to a milder power-law dependence on time once interactions start to compete with expansion, in agreement with previous results using different approximations~\cite{Berges:2013fga,Kurkela:2015qoa,Tanji:2017suk}. For comparison, we also show the result by taking only elastic collisions into account (dotted curve).
 The difference from the evolution with a full collision kernel is comparably small, indicating that the inelastic processes contribute mainly at low momenta, which are phase-space suppressed for bulk quantities such as pressure. 

\begin{figure*}[t!]
	\centering
	\includegraphics[scale=1]{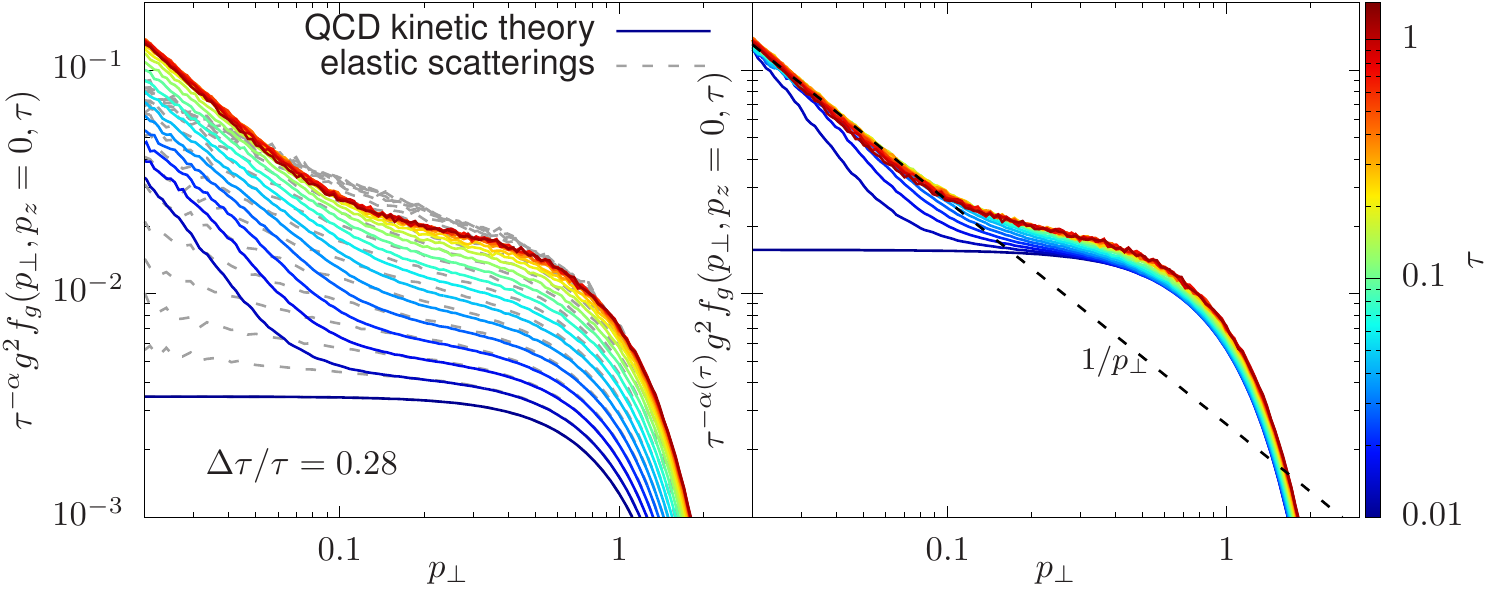}
	\caption{Left: Gluon distribution rescaled with $\tau^{-\alpha}$ versus transverse momentum for the full QCD collision kernels (solid) and with elastic scatterings only (dashed). Right: The same distribution, but rescaled with \emph{time-dependent} scaling exponent. 
}
	\label{fig:constscaling}
\end{figure*}

The emergence of scaling can be efficiently analyzed from moments of the distribution functions for gluons
\begin{equation}
n_{m,n}(\tau)\equiv\nu_g\int \frac{d^3\p}{(2\pi)^3}\, p_\perp^m |p_z|^ n f_g( p_\perp, p_z,\tau),\label{eq:moments}
\end{equation}
and equivalently for quarks. Effectively, different moments probe different momenta and are thus sensitive to scaling in a particular momentum regime.
If a distribution function shows prescaling (\ref{eq:prescaling}), the precise values of $\alpha(\tau), \beta(\tau)$ and $\gamma(\tau)$ in general depend on the history of the evolution from the reference time, which is normalized to one, and the final time $\tau$. Therefore we redefine the exponents in \eqref{eq:prescaling} to reflect the instantaneous scaling properties with 
\begin{equation} \tau^{\alpha(\tau)} \rightarrow \exp\left[\int_{1}^{\tau} \frac{d\tau' }{\tau' }\, \alpha(\tau')\right],\label{eq:timexp}
\end{equation}
which for constant $\alpha$ reduces to the power law $\tau^\alpha$. 
Then the rate of change of a particular moment $n_{m,n}$ is given by a linear combination of scaling exponents
\begin{equation}
\frac{d \log n_{m,n}(\tau)}{d \log \tau}=\alpha(\tau) -(m+2)\, \beta(\tau) -(n+1)\, \gamma(\tau).\label{eq:kexp}
\end{equation}
By looking at different moments $n,m=0,1,\ldots$ one obtains a set of algebraic equations from which $\alpha(\tau)$, $\beta(\tau)$ and $\gamma(\tau)$ can be determined.
Since the choice of moments is not unique, one can probe different momentum regimes and test how well \mbox{(pre-)scaling} is realized. 

The time-dependent exponents obtained from various combinations of moments $n_{n,m}$ with $n,m<4$ for initial conditions with $\sigma_0=0.1$ are shown in Fig.~\ref{fig:timeexp}, exhibiting a remarkable overlap of the results from different moment ratios~\footnote{We considered the following five triplets of moments:
$\{n_{0,0}, n_{1,0}, n_{0,1}\}$,
$\{n_{0,0}, n_{2,0}, n_{0,2}\}$,
$\{n_{1,0}, n_{2,0}, n_{1,1}\}$,
$\{n_{2,0}, n_{3,0}, n_{1,1}\}$,
$\{n_{0,0}, n_{3,0}, n_{0,3}\}$, form which sets of $\alpha(\tau),\beta(\tau),\gamma(\tau)$ were obtained according to \eqref{eq:kexp}.}.
In this case, one expects free-streaming scaling exponents $\alpha \rightarrow 0$, $\beta \rightarrow 0$, $ \gamma \rightarrow 1$ at very early times. Accordingly, both $\alpha(\tau)$ and $\gamma(\tau)$ approach the nonthermal fixed-point limit from above for initial conditions with $\sigma_0=0.1$~\footnote{Note that ideal hydrodynamics for massless particles shows scaling behavior with $\alpha \rightarrow 0$, $\beta \rightarrow 1/3$, $\gamma \rightarrow1/3$ for a local thermal distribution function.}.  
At later times $\tau>1$ we fit the power-laws with constant exponents and obtain $\alpha\approx-0.73$, $\beta\approx-0.01$ and $\gamma\approx0.29$. These values are close to both the analytic values of the BMSS and BD estimates given above and consistent with previous lattice results within errors~\cite{Berges:2013fga}.
One clearly observes the prescaling regime, for which different moments can be described by the common set of time-dependent scaling exponents even before the asymptotic scaling is reached. 
However, to emphasize that the time-dependence of the exponents is not universal, we show in \Fig{fig:plotexponentsqcddivv3b} the results for larger initial gluon density $\sigma_0=0.6$ such that free streaming is suppressed. In this case we see that at very early times $\tau<0.03$ there is no unique notion of scaling exponents. But very quickly the results from different sets of moments collapse again to a single curve, much before the exponents attain their universal constant values.

{\it Universal scaling form of the distributions.} With the results for exponents, we can now extract the universal scaling form $f_S$. We first consider rescaling with the constant values of exponents obtained from the late time fit.
The left panel of Fig.~\ref{fig:constscaling} shows the rescaled gluon distribution $\tau^{-\alpha} g^2 f_g$ as a function of $p_\perp$ at different times $\tau$ for $p_z = 0$ (solid lines) for initial conditions with $\sigma_0=0.1$. After an initial period, all rescaled curves at different times collapse to a single scaling curve. 

We see that with full collision kernel, the low-momentum part of the distribution function develops a $\sim 1/p_T$ behavior. In contrast, only elastic processes are not efficient in developing these thermal-like features of a low-momentum bath (grey dashed curves)~\footnote{This is in apparent contrast to elastic small-angle scattering approximations, where the $1/p_T$ behavior is also seen~\cite{Tanji:2017suk}.}. The softer momentum region is efficiently populated by the collinear radiation processes, and we observe excellent scaling properties also in that regime where particle number changing processes are essential. 

Prescaling states that the very same distribution function $f_S$ can be extracted at much earlier times, before the scaling exponents take on their universal values. To verify this, we rescale the distribution function according to \eqref{eq:prescaling} using the time dependent exponents from \Fig{fig:timeexp} and relation \eqref{eq:timexp}.  As shown in the right panel of  Fig.~\ref{fig:constscaling} the rescaled distribution collapses to a single scaling curve even at early times. As can be seen from Fig.~\ref{fig:PLPT}, the time-dependent exponents of Fig.~\ref{fig:timeexp} along with the universal scaling form $f_S$ can be established already at a time, where the bulk quantity $P_L/P_T$ still appears to be deep in the free-streaming regime.

\begin{figure}[t!]
	\centering
	\includegraphics[width=\columnwidth]{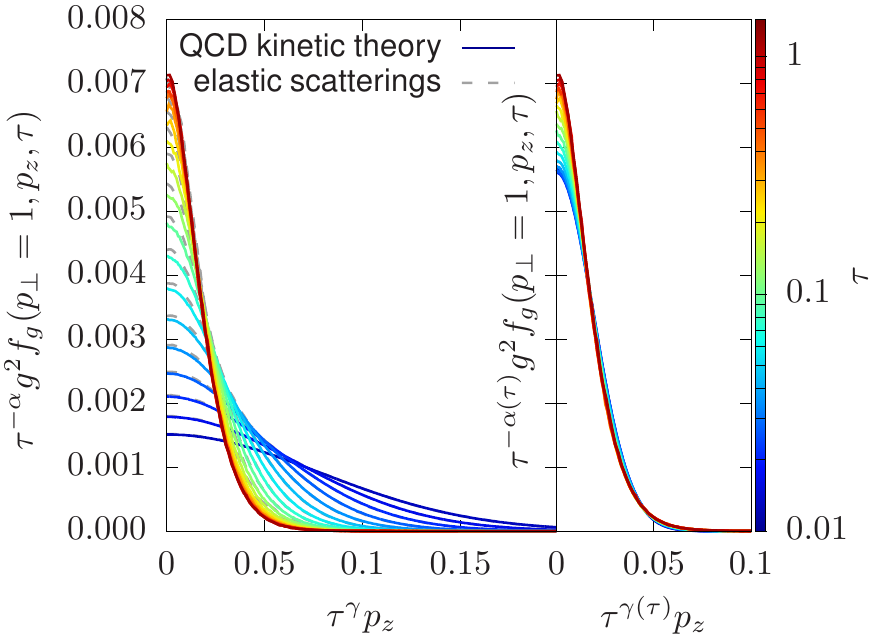}
	\caption{Left: Gluon distribution versus longitudinal momentum $\tau^{\gamma}p_z$. Right: Same but with \emph{time-dependent} exponents.}
	\label{fig:scalingpz}
\label{fig:prescalingpz}
\end{figure}

A corresponding analysis can be done for the longitudinal momentum dependence. Fig.~\ref{fig:scalingpz} displays the rescaled distribution as a function of $ \tau^{\gamma}p_z$ and $ \tau^{\gamma(\tau)}p_z$, respectively, at different times and we neglect the nearly vanishing transverse momentum exponent $\beta$. Again a much earlier collapse of the curves is observed if time-dependent exponents are used. 

\begin{figure}[t!]
	\centering
	\includegraphics[width=\columnwidth]{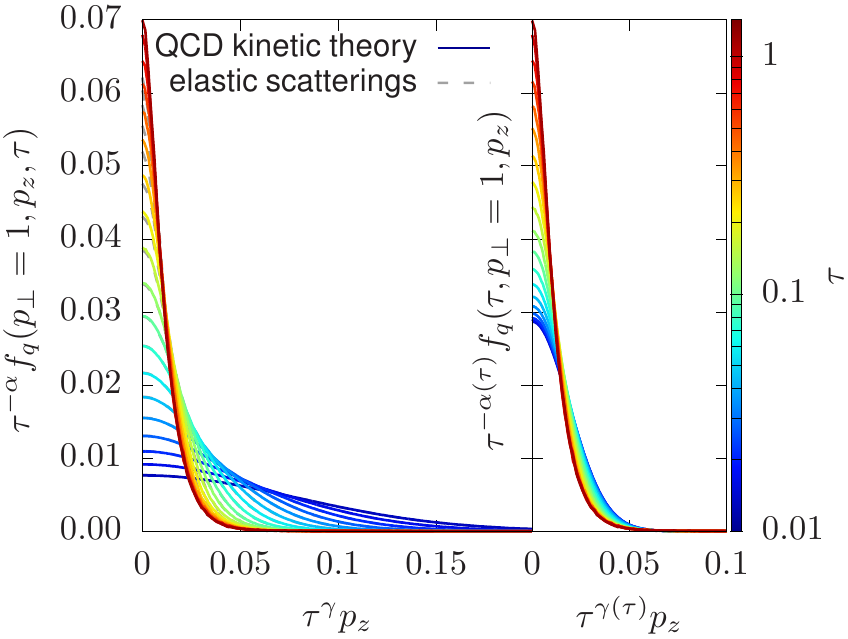}
	\caption{Left: Fermion distribution versus longitudinal momentum $\tau^{\gamma}p_z$. Right: Same but rescaled using the time-dependent gluon scaling exponents employed also in \Fig{fig:timeexp}.}
	\label{fig:fermionspz}
\end{figure}

Like for the case with small-angle scattering approximation~\cite{Tanji:2017suk}, we find that the quarks exhibit similar scaling behavior as for gluons at late times for the part of the distribution function not bounded by the Pauli exclusion principle. In \Fig{fig:fermionspz} we show the fermion distribution along the longitudinal momenta and $p_\perp/Q_s=1$. Although the time-dependent exponents capture most of the longitudinal squeeze of the distribution function, the scaling form of the fermion distribution function is not established as well as for gluons. Because gluons are highly occupied, the quark contribution to the total particle number is small at these times. Therefore, the background evolution of gluons does not change noticeably in the presence of quarks in this regime.


{\it Far-from-equilibrium hydrodynamic behavior.} It is remarkable to observe that the dynamics in the prescaling regime can be explained by the distribution function rescaling \eqref{eq:prescaling} with three slowly changing exponents $\alpha(\tau)$, $\beta(\tau)$ and $\gamma(\tau)$. 
This is precisely the situation one encounters in hydrodynamics, which is an effective description in terms of a few slowly varying degrees of freedom. To make this link more concrete, we 
consider the energy-momentum tensor $T^{\mu\nu}$ of (\ref{eq:enmomtensor}) along with the particle number current $J^\mu$ and a rank-three tensor $I^{\mu\nu\sigma}$, 
\begin{align}
J^{\mu} &=\nu_g \int \frac{d^3\p}{(2\pi)^3}\frac{p^\mu}{p^0} f_\p,\label{J}\\
I^{\mu \nu\sigma} &=\nu_g \int \frac{d^3\p}{(2\pi)^3}\frac{p^\mu p^\nu\label{I} p^\sigma}{p^0} f_\p ,
\end{align}
where we focus on the gluonic part, i.e.~$\nu_q = 0$. Integrating the kinetic equation (\ref{bolz}) with the appropriate powers of $p^\mu$ yields the equations of motion for these quantities.  

For the case of homogeneous boost-invariant expansion of a conformal system, \eqref{J} and \eqref{I} have only three independent components, namely the particle number density $n \equiv J^0$ and the tensors $2I^{0 xx}$ ($=2I^{0 yy}$) and $I^{0 zz}$, which together with $T^{00}$ evolve according to
 \begin{align}
&\partial_\tau n+ \frac{n}{\tau} = -C_J,\quad \partial_\tau T^{00}+ \frac{T^{00}+T^{zz}}{\tau} = 0 ,\\
&\partial_\tau I^{0 xx}+ \frac{I^{0 xx}}{\tau} = -C_I^{xx} ,\quad \partial_\tau I^{0 zz}+ \frac{3I^{0 zz}}{\tau} = -C_I^{zz} , \nonumber
\end{align} 
with the collision integrals $C_J= \nu_g\int d^3\p/((2\pi)^3 p^0) C[f]$ and $C_I^{\nu\sigma} = \nu_g\int d^3\p/((2\pi)^3 p^0) p^\nu p^\sigma C[f]$.
Noting that $n=n_{0,0}$, $2I^{0 xx}=n_{2,0}$ and $I^{0 zz}=n_{0,2}$, we see that  the equations of motion for these moments can be mapped to the same number of slowly varying scaling exponents using \eqref{eq:kexp}.
For a given distribution function  $f_S$, which is an input from the far-from-equilibrium QCD computation near the nonthermal attractor, the scaling dependence of the collision kernels $C_J$, $C_I^{xx}$ and $C_I^{zz}$ then closes the system of hydrodynamic equation of motions. 

What is special in comparison to more conventional hydrodynamics descriptions, which describe the evolution with respect to thermal equilibrium, is that our far-from-equilibrium hydrodynamics describes the evolution with respect to a nonthermal fixed-point distribution $f_S$.
In spirit this is similar to anisotropic hydrodynamic formulations, which are based on an expansion around a deformed equilibrium distribution~\cite{Romatschke:2003ms, Martinez:2012tu,Bazow:2013ifa}. 
However, our results establish the existence of a new hydrodynamic regime far away from equilibrium at early times, which has not much to say about the later time approach to thermal equilibrium, where more conventional hydrodynamic descriptions should be applied. The subsequent evolution of the quark-gluon plasma towards thermal equilibrium with QCD kinetic theory is the subject of a separate work~\cite{Kurkela:2018oqw, Kurkela:2018xxd}. 

{\it Conclusion.} Our study presents the full solution of the leading-order QCD kinetic equations with quarks and gluons  in the nonequilibrium regime. The results demonstrate the emergence of early hydrodynamic behavior around a far-from-equilibrium state, which is qualitatively different from the more conventional hydrodynamics around equilibrium. The effective description at early times is based on three slowly varying degrees of freedom---the time dependent scaling exponents. The exponents relax to the constant values characterizing the nonthermal fixed point, whose existence we confirm in the presence of both elastic and particle number changing processes. However, it is a particular strength of our findings that they apply to systems which are still away from the asymptotic scaling regime of a nonthermal fixed point. Therefore even in cases where scaling is never reached, the evolution may be described by prescaling dynamics.

More generally, our work provides new insights into the important question of memory loss at early stages in complex systems far from equilibrium and the establishment of effective theories like hydrodynamics from the underlying microscopic physics. Though we have focused on prescaling in the quark-gluon plasma, 
the generalized prescaling relation \eqref{eq:prescaling} should be 
 relevant to other far-from-equilibrium many-body systems. For instance, earlier theoretical~\cite{Orioli:2015dxa,Berges:2015ixa,Mikheev:2018adp} and very recently also experimental~\cite{Prufer:2018hto,Erne:2018gmz} studies on quenches in nonequilibrium Bose gases may be interpreted along these lines, and it would be very interesting to revisit the results in view of our findings.

{\it Acknowledgments.} We thank K.~Boguslavski, G.S.~Denicol, S.~Erne, T.~Gasenzer, A.~Kurkela, A.~Mikheev, J.~Noronha, S.~Schlichting, J.~Schmiedmayer, N.~Tanji, R.~Venugopalan, and C.~Wetterich for discussions. This work is part of and supported by the DFG Collaborative Research Centre ``SFB 1225 (ISOQUANT)''.

\bibliography{master.bib}

\end{document}